\documentclass[manuscript]{aastex62}

\usepackage{CJK}
\graphicspath{{./}{figures/}}

\received{--}
\revised{--}
\accepted{--}
\submitjournal{ApJS}

\shorttitle{Open cluster within 500~pc}
\shortauthors{Qin et al.}

\begin{document}
\begin{CJK*}{UTF8}{gbsn}
\title{Hunting for Neighboring Open Clusters with {\it Gaia} DR3: \\101 New Open Clusters within 500~pc}

\correspondingauthor{Songmei Qin, Li Chen}
\email{qinsongmei@shao.ac.cn, chenli@shao.ac.cn}

\author[0000-0003-3713-2640]{Songmei, Qin (秦松梅)}
\affiliation{Key Laboratory for Research in Galaxies and Cosmology, Shanghai Astronomical Observatory, Chinese Academy of Sciences,80 Nandan Road, Shanghai 200030, China}
\affiliation{School of Astronomy and Space Science, University of Chinese Academy of Sciences, No. 19A, Yuquan Road, Beijing 100049, China}

\author[0000-0001-5245-0335]{Jing, Zhong (钟靖)}
\affiliation{Key Laboratory for Research in Galaxies and Cosmology, Shanghai Astronomical Observatory, Chinese Academy of Sciences,80 Nandan Road, Shanghai 200030, China}

\author[0000-0003-1864-8721]{Tong, Tang (唐通)}
\affiliation{Key Laboratory for Research in Galaxies and Cosmology, Shanghai Astronomical Observatory, Chinese Academy of Sciences,80 Nandan Road, Shanghai 200030, China}
\affiliation{School of Astronomy and Space Science, University of Chinese Academy of Sciences, No. 19A, Yuquan Road, Beijing 100049, China}

\author[0000-0002-4907-9720]{Li, Chen (陈力)}
\affiliation{Key Laboratory for Research in Galaxies and Cosmology, Shanghai Astronomical Observatory, Chinese Academy of Sciences,80 Nandan Road, Shanghai 200030, China}
\affiliation{School of Astronomy and Space Science, University of Chinese Academy of Sciences, No. 19A, Yuquan Road, Beijing 100049, China}

\begin{abstract}

We systematically searched for open clusters in the solar neighborhood within 500~pc using pyUPMASK and HDBSCAN clustering algorithms based on {\it Gaia} DR3. Taking into consideration that the physical size for most open clusters is less than 50~pc, we adopted a slicing approach for different distance shells and identified 324 neighboring open clusters, including 223 reported open clusters and 101 newly discovered open clusters (named as OCSN, Open Cluster of Solar Neighborhood). Our discovery has increased the number of open clusters in the solar neighborhood by about 45\%. In this work, larger spatial extents and more member stars were attained for our cluster sample. We provided the member stars and the membership probabilities through the pyUPMASK algorithm for each cluster and derived their astrophysical, age, and structural parameters.

\end{abstract}


\section{Introduction} 
\label{sec:intro}

Open clusters (OCs) in the Milky Way are a collection of stars that are formed from the same molecular cloud and gravitationally bound together \citep{2003ARA&A..41...57L,2010ARA&A..48..431P}, thus sharing similar specific characteristics (e.g. age, distance, reddening, and metal abundance, etc.). OCs provide an ideal laboratory for studying star formation and evolution \citep{2009ApJS..181..321E}. Meanwhile, with its large sample, OCs are powerful tracers of the Galactic disk to constrain the Galactic structure and evolution history \citep{1982ApJS...49..425J,2005ApJ...629..825D}.

As most Galactic OCs are located on the thin disc \citep{2013A&A...558A..53K}, observations for OCs are often hampered by the contamination from dominant background/foreground field stars, leading to more uncertainties in the characterization of cluster properties. Before the Gaia era, reliable member star selection was fairly difficult due to the limited astrometric precision, which brought about inconsistency in the determination of basic parameters like distance, kinematics, and age of OCs \citep{2015A&A...582A..19N}. 

As one of the most successful and ambitious projects, {\it Gaia} enabled a deluge of scientific work including improving the quality of the cluster census. The {\it Gaia} DR2 catalog \citep{2018A&A...616A...1G} presents more than 1.3 billion stars with unprecedented high-precision astrometric and photometric data, greatly improving the reliability of stellar membership determination and characterization of a large sample of OCs. The study of Galactic OCs has ushered in a new era. A more solid assessment of the cluster membership using the precise {\it Gaia} data greatly promoted subsequent discoveries of new OCs. In the meantime, with the popularity of machine learning, cluster census efficiency has been tremendously raised. Many different methods have been used to search for new or re-detect existing OCs with {\it Gaia} data, especially various clustering algorithms such as unsupervised photometric membership assignment in stellar clusters (UPMASK, \citet{2014A&A...561A..57K}), Density-Based Spatial Clustering of Applications with Noise (DBSCAN, \citet{1996A}), Hierarchical Density-Based Spatial Clustering of Applications with Noise (HDBSCAN, \citet{2013Density,mcinnes_hdbscan_2017}), Gaussian Mixture Models(GMMs, \citet{1894RSPTA.185...71P}), Friends of Friends Algorithm (FOF, \citet{1982ApJ...257..423H}), etc. \citet{2021A&A...646A.104H} have made detailed comparisons for three main algorithms (DBSCAN, HDBSCAN, GMMs) side-by-side, exploring their effectiveness in blind-searching of OCs with large-scale {\it Gaia} dataset. 

Based on {\it Gaia} DR2, \citet{2018A&A...618A..93C} applied the UPMASK algorithm to select cluster members and provided an updated catalog of 1229 OCs including previously reported clusters and 60 newly discovered clusters. \citet{2019ApJS..245...32L} reported 76 new OCs by employing a FOF-based cluster finder method to hunt out overdensities in the ($l$, $b$, $\varpi$, $\mu_{\alpha}^*$ \footnote{$\mu_{\alpha}^*=\mu_{\alpha}cos\delta$}, $\mu_{\delta}$) space. In parallel, using the DBSCAN algorithm, \citet{2020A&A...635A..45C} found 582 new cluster candidates located in the low galactic latitude area by investigating the aggregation of stars in the 5-dimensional parameter phase space, with part of the candidates overlapped with the result of \citeauthor{2019ApJS..245...32L}. Later on, \citet{2020A&A...640A...1C} compiled a comprehensive list of 1867 bona fide OCs reanalyzed with UPMASK and {\it Gaia} data, providing a large and homogeneous catalog of open cluster properties and the corresponding member stars. 

As the early stage of {\it Gaia} third data release, {\it Gaia} EDR3 \citep{2021A&A...649A...1G} provides astrometric and photometric parameters of 1.5 billion sources with even higher accuracy than {\it Gaia} DR2, averagely increasing the precision of proper motion by 2-3 times and parallax by about 20\%. Solely based on the Gaia EDR3 database, \citet{2022A&A...661A.118C} have proposed an OCfinder method, which employs a DBSCAN clustering algorithm for selecting over-densities in the five-dimensional astrometric space and incorporates a deep artificial neural network for distinguishing bona fide OCs in the color-magnitude diagrams (CMDs). In their work, 628 new OCs were picked out, mostly located 1~kpc further away from the Sun. In our recent work \citep{2022ApJS..260....8H}, we carried out a blind search for new OCs using {\it Gaia} DR2/EDR3 by dividing low galactic latitude regions into $2^{\circ} \times 2^{\circ}$ ($|b| < 5^{\circ}$) or $3^{\circ} \times 3^{\circ}$($|b| > 5^{\circ}$) grids, sequentially performing the DBSCAN and pyUPMASK clustering algorithms in five-dimensional phase space ($d_{l^*}$, $d_b$, $v_{\alpha^*}$, $v_{\delta}$, $\varpi$). Eventually, 541 star clusters unrecorded in literature were found, with a majority of them located beyond 1~kpc from the Sun, while only about 6\% are located within 500~pc to the Sun. Using a similar approach and with {\it Gaia} EDR3, \citet{2022ApJS..262....7H} expanded their search grid to $12^{\circ} \times 12^{\circ}$ or $18^{\circ} \times 18^{\circ}$ to increase the number of clusters in the solar vicinity and cataloged 270 newly found cluster candidates within 1.2~kpc of the Sun, of which 179 new OC candidates are within 500~pc.

In most cases of the above systematic searches for new OCs, {\it Gaia}'s high-precision position ($l$, $b$), proper motion ($\mu_{\alpha}^*$, $\mu_{\delta}$) and parallax ($\varpi$) data were used to detect the aggregation characteristics of cluster members in multi-dimensional phase space, photometric data were assisted in confirming the reality of the OCs in the CMD and determining their basic properties by subsequent further analysis. On the other hand, these works have different features on their own methodology and procedure. The newly found objects greatly enrich our understanding of the Galactic OCs population, and they also indicate that the present OCs sample is far from complete. It is anticipated that many new OCs still can be detected through careful analysis of observational data.

The main difficulty in detecting OCs lies in the serious contamination of field stars, especially for nearby OCs, in which member stars cover a wide projected field of view and the proportion of background field stars could be overwhelming. This study aims to find new OCs in the solar neighborhood, with an effective way to decontaminate the background/foreground objects by astrometric data. For this purpose, we proposed a new approach of data ``slicing" along the line-of-sight distance. In a systematic search of nearby OCs, we divided all-sky data into sub-blocks with different sizes in every 100~pc interval for reducing contamination of field stars as much as possible and highlighting the aggregation signal of cluster members in the multi-dimensional phase space. 

This paper is the first of our serial work, focused on a comprehensive search for undetected clusters within 500~pc from the Sun, solely based on the most recent precise {\it Gaia} DR3. With our novel “Slicing” approach, we greatly improved the efficiency of detecting the nearby new OCs. Altogether we updated/derived the physical properties of 324 OCs, with 223 previously reported clusters and 101 newly detected clusters. The data we used in this work is introduced in Section~\ref{sec: data}. In Section~\ref{sec: method}, we describe the searching process with the new approach of data ``slicing" along the line-of-sight distance. In Section~\ref{sec: catalog}, we describe the catalogs of OCs and their members. The discussion about cluster cross-match and property analysis is given in Section~\ref{sec: discussion}. Finally, we make a summary in Section~\ref{sec: sum}.

\section{Data} 
\label{sec: data}

This study aims to find new OCs in the solar neighborhood, with an efficient way of decontaminating the background influences. For this purpose, we restricted our survey to the stars within 500~pc, making full use of the high precision of {\it Gaia} DR3 data. 

{\it Gaia} Data Release 3 ({\it Gaia} DR3; \citet{2022arXiv220800211G}) provides astrometric information with nearly the same high-precision as {\it Gaia} EDR3 for about 1.8 billion sources over the sky, as well as near-millimagnitude precision photometric data in three bands ($G$, $G_{\rm BP}$ and $G_{\rm RP}$). Moreover, {\it Gaia} DR3 introduces an impressive wealth of new data products, including the accurate radial velocity parameters for more than 33 million objects \citep{2022arXiv220605902K} and atmospheric parameters ($T_{\rm{eff}}$, log$g$ and [M/H]) for about 470 million sources \citep{2022arXiv220605992F}. For {\it Gaia} DR3, the typical proper motion uncertainty respectively goes from 0.07 mas yr$^{-1}$ for $G \approx$ 17~mag, up to 0.5 mas yr$^{-1}$ for $G =$ 20~mag, the parallax uncertainty goes from 0.07~mas at $G \approx$ 17~mag, up to 0.5~mas at $G =$ 20~mag, and the mean G-band photometry uncertainty goes from 1~mmag at $G \approx$ 17~mag, up to 6~mmag at $G =$ 20~mag.  It is noteworthy that the newly determined median radial velocities \citep{2022arXiv220605902K} have risen greatly in number compared to {\it Gaia} DR2 and its median precision goes from 1.3~km s$^{-1}$ at $G_{\rm RVS} \approx$ 12~mag, up to 6.4~km s$^{-1}$ at $G_{\rm RVS} =$ 14~mag, which greatly help us to study the dynamical evolution of the OCs in the Milky Way. 

In this work, we first select stars from {\it Gaia} DR3 in the region with $|b| < 30^{\circ}$ since the majority of identified OCs are located in $|b| < 20^{\circ}$ \citep{2002A&A...389..871D,2013A&A...558A..53K,2020A&A...635A..45C}. Then, we used the \textbf{astroquery.gaia} python package\footnote{https://astroquery.readthedocs.io/en/latest/gaia/gaia.html} \citep{2019AJ....157...98G} to obtain the positions ($l$, $b$), proper motions ($\mu_{\alpha}^*$, $\mu_{\delta}$), parallaxes ($\varpi$), magnitudes in three photometric filters ($G$, $G_{\rm BP}$ and $G_{\rm RP}$), radial velocity ($rv$), and their associated uncertainties from {\it Gaia} Archive\footnote{https://gea.esac.esa.int/archive/}. To investigate the OCs in the solar vicinity, we only retained stars with parallax greater than 2, approximately corresponding to a distance of 500~pc. To reduce the faint background stars, we filtered out stars with $G > 18$~mag. And we applied the cut on re-normalized unit weight error (RUWE) $<$ 1.4 \citep{LL:LL-124} to exclude the unreliable astrometric and photometric data. Finally, we screened out about 8 million stars as the initial sample.

\section{Method}
\label{sec: method}

Our search process is divided into four steps:
\begin{enumerate}
\item [(1)] Partitioning the entire sample into multi-blocks to reduce the contamination of field stars， see details in Section~\ref{sec: cut}.
\item [(2)] Employing an unsupervised clustering method pyUPMASK in 5-dimensional data ($l$, $b$, $\varpi$, $\mu_{\alpha}^*$, $\mu_{\delta}$) to assign probabilities for the stars and remove the stars with probabilities $<$ 0.1 in each block. Then we pick out those blocks that may contain OCs from proper motion distribution features, see details in Section~\ref{sec: pyupmask}.
\item [(3)] Separating the stars in each block into members of individual cluster candidates with HDBSCAN, see details in Section~\ref{sec: hdbscan}.
\item [(4)] By applying visual inspection on the distribution of position ($l$, $b$), proper motion ($\mu_{\alpha}^*$,$\mu_{\delta}$), magnitude-parallax ($G$, $\varpi$), and color-magnitude diagram ($G_{\rm BP}-G_{\rm RP}$, $G$) of each cluster candidate, we confirm the true existence of the star clusters, see details in Section~\ref{sec: checking cmd}.
\end{enumerate}

According to the parameter ranges obtained in the above identification process, we retrieved the astrometric and photometric data for stars in the cluster regions. Then, we obtained the membership probabilities of stars in each cluster through the pyUPMASK algorithm and regarded the stars with probabilities greater than 0.5 as cluster members, which are cataloged in Table~\ref{member cat}.

\subsection{Data slicing}
\label{sec: cut}

\begin{figure*}[h]
   \centering
   \includegraphics[width=\textwidth, angle=0]{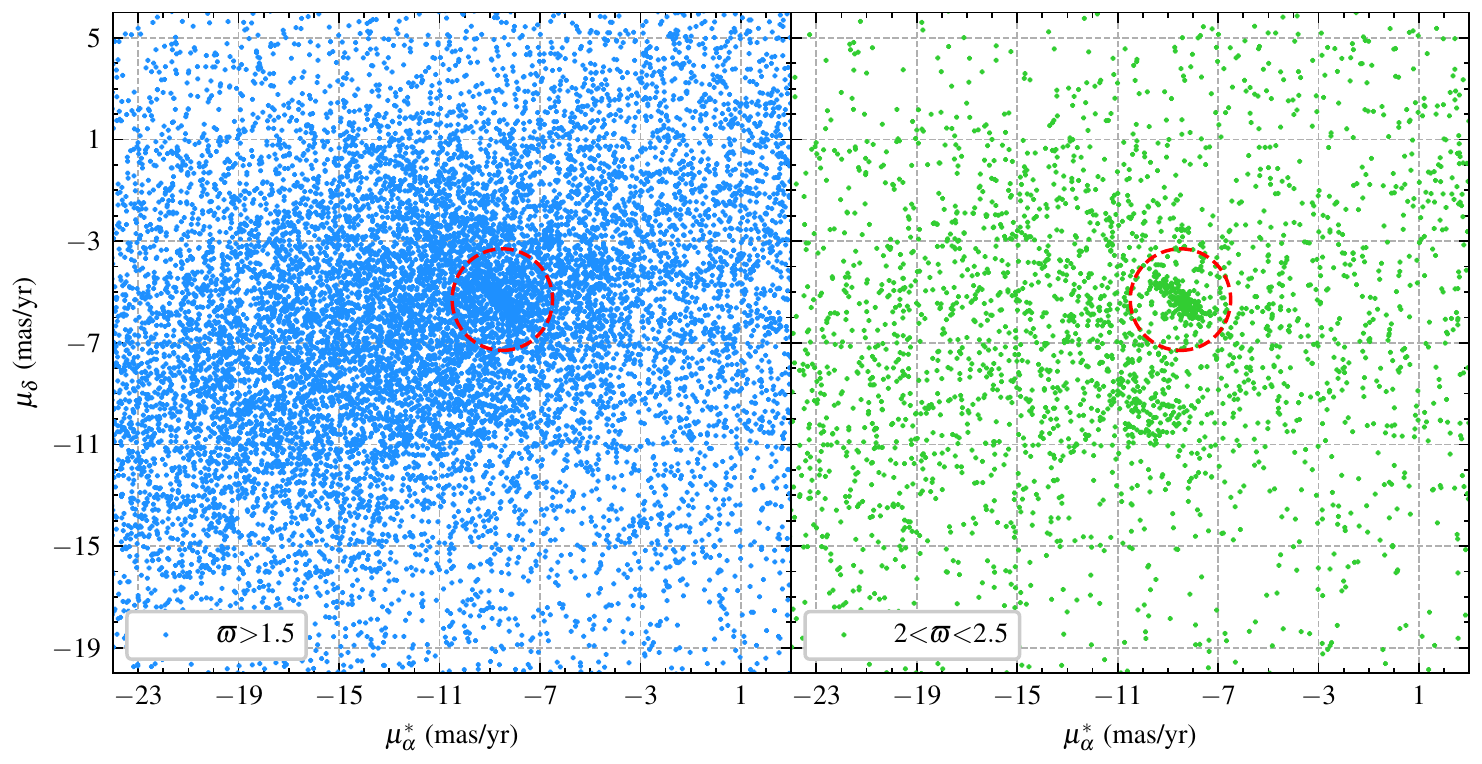}
   \caption{The proper motion distribution of two samples. The blue and green dots refer to stars with different distant slicing. The red dashed circles represent the OC where the members assembled.}  
   \label{fig1}
\end{figure*}

Members of an open cluster that formed in the same molecular clouds, generally are severely immersed in a dense star field and it is hard to distinguish the actual star cluster members from the background, especially for those clusters mixed with the vast majority of field stars. The most efficient way of hunting out new clusters is to look for the clustering of stars in the velocity space (i.e. the vector-point-diagram) since open cluster members share a distinctive movement as compared to the field stars. For highlighting the cluster members in the proper motion distribution diagram, we cut the entire sample into various sub-samples according to their 3D spatial coordinates ($l, b$, and $\varpi$). For example, as we can see in Figure~\ref{fig1}, the two panels are the proper motion distributions ($\mu_{\alpha}^*$, $\mu_{\delta}$) of two sub-samples with the same 2D spatial distribution range: $306^{\circ}$ $\le$ $l$ $\le$ $312^{\circ}$ and $-6^{\circ}$ $\le$ $b$ $\le$ $0^{\circ}$. The $\varpi$ of the left sample is greater than 1.5, in the red dashed circle we can see a vaguely condensed area blurred by the overwhelming distribution of field stars. In the right panel of Figure~\ref{fig1}, we limited this sample to $2 < \varpi < 2.5$, thus removing many foregrounds and background stars and the corresponding distribution shows an obvious concentration of cluster member stars around the same region. Hence, ingenious data slicing is critical for hunting out new OCs in our cluster searching volume. 

To  more fully and efficiently reveal OCs hidden in the field stars, we divided the entire remaining stellar sample within the Galactic latitude range of $|b| < 30^{\circ}$ into 1044 blocks according to their 3D spatial coordinates ($l$, $b$, and $\varpi$). Firstly, We separated all the stars into several searching shell regions along the $\varpi$ ($>10$~mas, 5-10~mas, 3.33-5~mas, 2.5-3,33~mas, 2-2.5~mas), approximately corresponding to the distance ($<100$~pc, 100-200~pc, 200-300~pc, 300-400~pc, 400-500~pc), which is large enough to cover at least one typical open cluster. In recent years, many investigations based on {\it Gaia} data revealed that besides the high-density inner parts, there might be extended low-density outer halos \citep{2019A&A...624A..34Z,2021A&A...645A..84M,2022AJ....164...54Z} or elongated tidal tails \citep{2019A&A...627A.119C,2020ApJ...889...99Z,2022RAA....22e5022B,2022MNRAS.514.3579B} in the outskirt of the OCs. To adequately contain the outer halo structure of OCs, we adopted 50~pc as the typical extended spatial scale of OCs \citep{2022A&A...659A..59T,2022AJ....164...54Z}. As the projected angular size varies with distance, we adopted the searching grids with different angular sizes, ranging from $30^{\circ}$ at a close distance of 100~pc to about $6^{\circ}$ at 500~pc, as listed in Table~\ref{tab1}.

\begin{deluxetable}{cccccc}[h]
\caption{Size of the search area. \label{tab1}}
\tablewidth{1pt}
\tablehead{\colhead{dist (pc)} & \colhead{$<$100} & \colhead{100-200} &
\colhead{200-300} & \colhead{300-400} & \colhead{400-500}\\
$\varpi$ (mas) & $>$10 & 5-10 & 3.33-5 &  2.5-3.33 & 2-2.5
}
\startdata
 size (deg) & 30 & 20 & 12 & 10 & 6\\
\enddata
\end{deluxetable}

\subsection{Initial screening with pyUPMASK}
\label{sec: pyupmask}

After getting the 1044 slicing blocks, we applied the pyUPMASK algorithm \citep{2021A&A...650A.109P} to gain membership probabilities of each star, which is based on the clustering of members compared to field stars in ($l$, $b$, $\varpi$, $\mu_{\alpha}^*$, $\mu_{\delta}$) phase space. pyUPMASK is an open-source software package compiled by Python language following the development principle of UPMASK \citep{2014A&A...561A..57K}, which is a member star determination method developed to process photometric data originally, though later widely used in the determination of member stars based on astrometric data \citep{2018A&A...618A..93C,2020A&A...640A...1C}. 

This enhanced clustering method contains several major procedures as follows: ($\textrm{i}$) Input data reduction using Principal Component Analysis (PCA); ($\textrm{ii}$) Choose one of the clustering algorithms such as K-means\citep{2001K}, mini-batch k-means (MBK, \citet{2010Web}), Gaussian mixture models (GMM, \citet{1894RSPTA.185...71P}), agglomerative clustering (AGG, \citet{2013Hierarchical}), the nearest neighbors density method (KNN, \citet{2014Sci...344.1492R}), Voronoi (VOR, \citet{0Nouvelles}) method supported by pyUPMASK to process the reduced data; ($\textrm{iii}$) Employ Ripley's K function \citep{1976The} to assess the authenticity of the clusters (or reject the fake clusters with a random uniform distribution); ($\textrm{iv}$) Apply the Gaussian-Uniform Mixture model to level down the field contamination. We skip this step to reserve those poor or non-Gaussian distribution clusters; ($\textrm{v}$) Evaluate the cluster membership probabilities through the kernel density estimator (KDE). 

\begin{figure*}[h]
   \centering
   \includegraphics[width=\textwidth, angle=0]{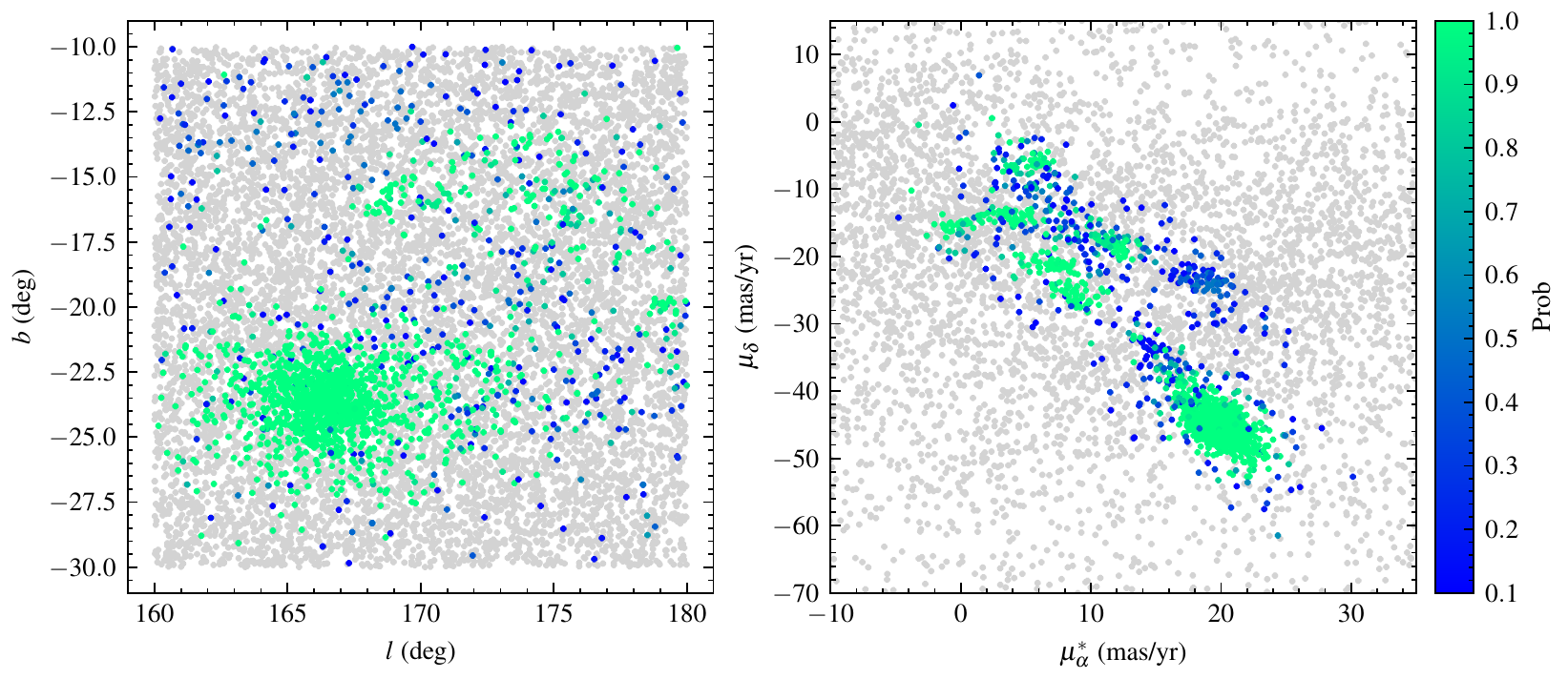}
   \caption{Position and proper motion distributions of one block sample. The grey dots refer to the stars with membership probabilities less than 0.1, while the colored dots are stars with membership probabilities greater than 0.1. The color bar represents the probabilities of member stars.}  
   \label{po-pm}
\end{figure*}

The continuous KDE probabilities between 0 and 1 are assigned to all the stars defined as $P_{cl} = KDE_{m} / (KDE_{m} + KDE_{nm})$, where $KDE_{m}$ and $KDE_{nm}$ refer to the KDE likelihoods for the members and nonmembers, and this reflects the proportion of true cluster stars and field stars. Concerning the imbalance of members and field stars that are mixed within a stellar cluster field, we set the lower limit of member probabilities to 0.1 to preserve cluster members as many as possible. As an example, Figure~\ref{po-pm} shows the position and proper motion distributions of a sample with the range of $160^{\circ}$ $\le$ $l$ $\le$ $180^{\circ}$, $-30^{\circ}$ $\le$ $b$ $\le$ $-10^{\circ}$ and $5~mas < \varpi < 10~mas$. In Figure~\ref{po-pm}, grey dots represent stars with probabilities less than 0.1 (about 90\%), while the rest colored dots represent stars with probabilities greater than 0.1 (about 10\%). We can see that after rejecting a large proportion of field stars, some clumps are exposed in the position and proper motion distributions. Nevertheless, some clusters such as the Cand1 in Figure~\ref{5-dim} show a relatively sparse spatial distribution that leads to lower probabilities. To avoid losing such sparse OCs, we applied the cut at probability $> 0.1$ to exclude most field stars. 

After excluding field stars with low probability (p$< 0.1$) in all block samples, we visually inspected the proper motion distributions of member stars. As shown in Figure~\ref{po-pm}, the block sample exhibits several distinct clumps both in position and proper motion distribution, which would be reserved in the following analysis in Section~\ref{sec: hdbscan}. It is noticed that, the concentrated structure around (20~$mas yr^{-1}$, $-$45~$mas yr^{-1}$) is the famous open cluster Pleiades (M 45) \citep{2020A&A...640A...1C,2022ApJ...926..132H}. Eventually, 394 block samples that might contain OCs were retained.

\subsection{Separating OCs with HDBSCAN}
\label{sec: hdbscan}

For the 394 samples, we normalized the data in 5 dimensions($l$, $b$, $\varpi$, $\mu_{\alpha}^*$, $\mu_{\delta}$) and then used the HDBSCAN \citep{mcinnes2017hdbscan} to separate individual cluster candidate members from field stars. HDBSCAN is primarily  proposed by \citeauthor{2013Density}, which combines the density-based approach of DBSCAN \citep{1996A} with hierarchical clustering to deal with datasets of varying densities. The key parameter to affect the resulting clustering is $min\_cluster\_size$ ($m_{clSize}$), which refers to the minimum possible size of a cluster. The smaller value of $m_{clSize}$ might cause a large cluster to be divided into small clumps, which may generate some fake clusters. However, with a larger value of $m_{clSize}$, small adjacent clusters in phase space would be combined as a big one.

\begin{figure*}[h]
   \centering
   \includegraphics[width=0.7\textwidth, angle=0]{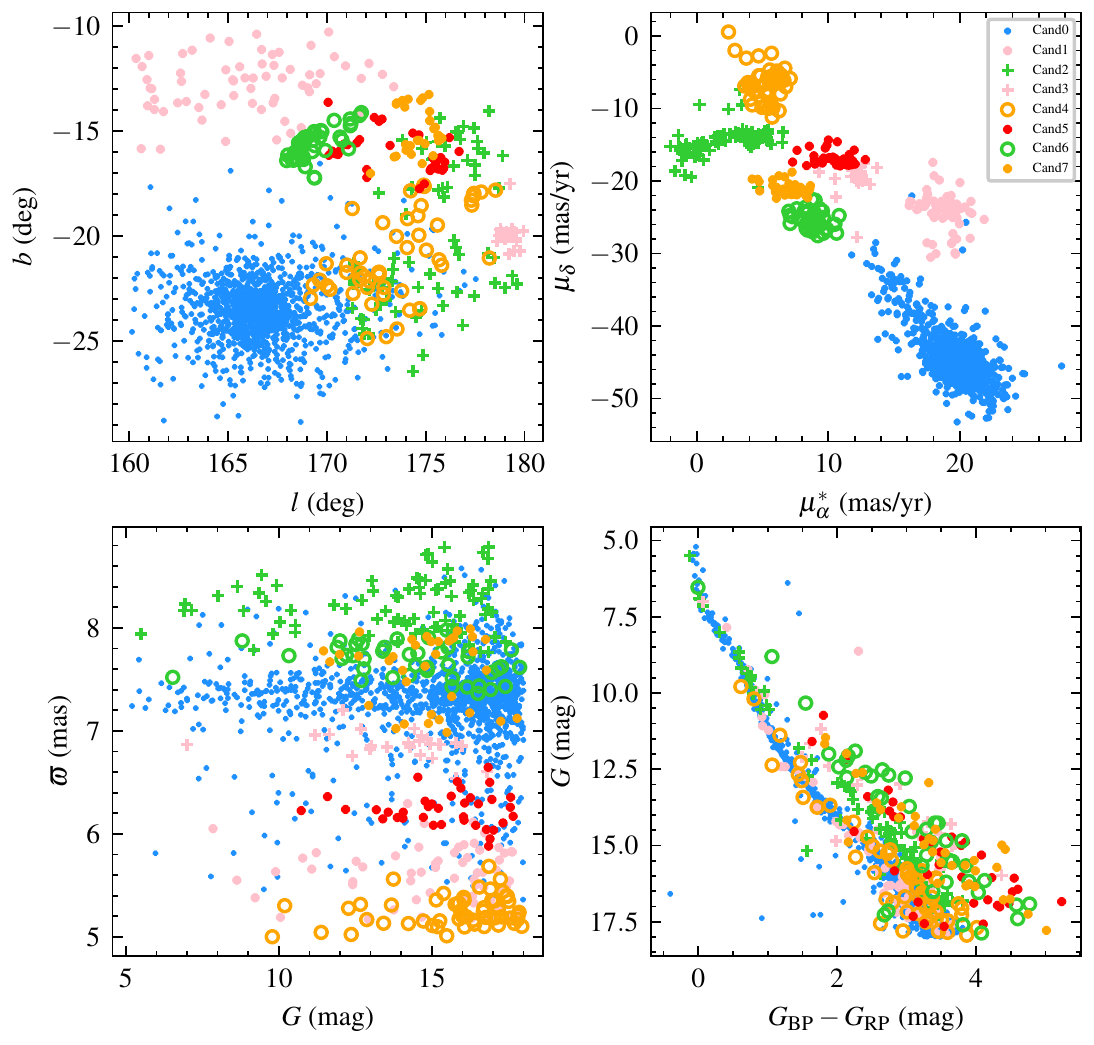}
   \caption{Spatial ($l$, $b$), proper motion ($\mu_{\alpha}^*$, $\mu_{\delta}$), magnitude-parallax ($G$, $\varpi$), color-magnitude ($G_{\rm BP}-G_{\rm RP}$, $G$) diagrams of the eight separated cluster candidates using HDBSCAN. Colors and symbols represent member stars of different cluster candidates.}  
   \label{5-dim}
\end{figure*}

\citet{2021A&A...646A.104H} compared three clustering algorithms DBSCAN, HDBSCAN and GMMs in computational speed and availability, and concluded that HDBSCAN is the most sensitive and effective method for revealing OCs in {\it Gaia} data. In their work, the performance of the HDBSCAN algorithm for 100 OCs showed that the value of ``Sensitivity" = TP / (TP + FN) was the largest when adopted $m_{clSize}$ as 10, which corresponds to the strongest ability of HDBSCAN to detect real OCs. \citet{2013Density} also recommended setting $m_{clSize} = m_{Pts} = 10$ for best sensitivity and speed when running the algorithm. Thus, we assigned $m_{clSize} = 10$. At the same time, for better detecting some sparse OCs, we selected the ``leaf'' cluster selection method \citep{mcinnes2017hdbscan}. After applying HDBSCAN to separate out cluster groups in 5-dimension data, we obtained 800 OC candidates. For example, in Figure~\ref{5-dim}, the same sample stars as in Figure~\ref{po-pm} were separated into eight cluster candidates with HDBSCAN. Although some fake clusters may arise because of the parameter selection in HDBSCAN, we will check every cluster candidate in 5-dimension data by eyes, see details in Section~\ref{sec: checking cmd}.

\subsection{Visual inspection}
\label{sec: checking cmd}

The member stars in an OC are co-moving and sharing similar parallax/distance. In the meantime, the color-magnitude diagram of the members is expected to present a clear main-sequence feature. We visually screen each cluster candidate in terms of position ($l$, $b$), proper motion ($\mu_{\alpha}^*$, $\mu_{\delta}$), magnitude-parallax ($G$, $\varpi$) distributions and color-magnitude diagram ($G_{\rm BP}-G_{\rm RP}$, $G$) to reject the ``false positive'' clusters. 

\begin{figure*}[h]
   \centering
   \includegraphics[width=\textwidth, angle=0]{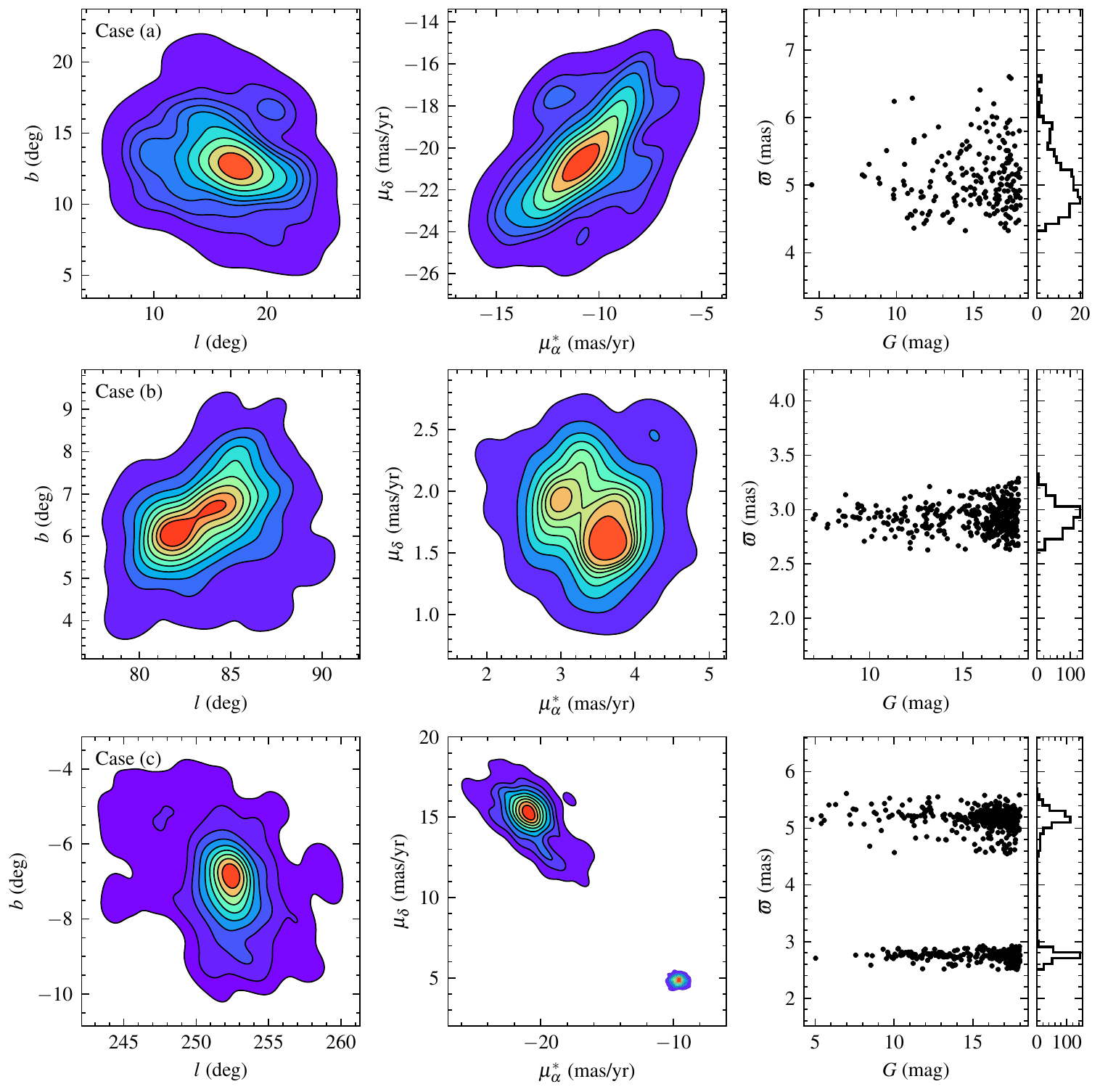}
   \caption{Three examples of target and reference clusters. The left panels present the spatial distributions of members' iso-density contours. The middle panels present the proper motion distribution of members' iso-density contours. The right panels present the $G$-parallax distribution with parallax histograms on the right edge.}  
   \label{merge cases}
\end{figure*}

It is noticed that a given cluster candidate may happen to be identified within one block, sometimes may fall on the borders, or can be detected in more than one block (with different membership results). To merge the split cluster candidates, we adopted the following process. We designated a nearby cluster with the smallest angular distance as a reference cluster for each target cluster identified by eye inspection. And we inspected the position ($l$, $b$), proper motion ($\mu_{\alpha}^*$, $\mu_{\delta}$), magnitude-parallax ($G$, $\varpi$) distribution and color-magnitude diagram ($G_{\rm BP}-G_{\rm RP}$, $G$) of the target and reference clusters together to assess whether they are the same cluster or not. A few specific cases that arose in our cluster samples are listed as follows: 
\begin{enumerate}
\item [(a)] Only one peak appears in both spatial and proper motion distributions. If the parallaxes of their members are consistent, the two clusters are thought to be combined together; if not, the two clusters are considered to be different individuals.
\item [(b)] More than one peak arises in both spatial and proper motion distributions. In this case, the two clusters are regarded as different individuals. 
\item [(c)] Only one peak in the spatial distribution but two separated peaks in the parallax histogram. They are identified as distinguished clusters. 
\end{enumerate}
Figure~\ref{merge cases} shows 5-dimensional distributions of three corresponding examples. The example of the case ($\textrm{a}$) is presented in the top panel: only one spatial over-density distribution can be detected and one peak in the proper motion distribution. In addition, there is one peak around 5~mas in the parallax histogram. These properties demonstrate that all members belong to the same cluster. The middle panels display the case ($\textrm{b}$) example of two clusters, which hold close but notably different peaks. And their mean parallaxes are approximately equal to 2.9~mas. They could be binary clusters at nearly the same distance. To illustrate the case ($\textrm{c}$), we show two distinct clusters which have the same projected position in the bottom panel. Although these clusters overlap in the projected position, which means one is in the front and the other is in the back, they can be well distinguished through the distribution of different proper motions and parallaxes. During the above visual checking process, we adjusted the parameter ranges of each cluster. Finally, we obtained 324 genuine stellar aggregations, most of which are certainly OCs. 

\section{Cluster and Member Catalogue}
\label{sec: catalog}

We regarded stars with membership probabilities $>$ 0.5 as cluster members. Based on those members, we attained the cluster properties. We provided two catalogs\footnote{The complete catalogs are available on CDS.} in this paper: one for the properties of 324 OCs and the other for parameters of 59304 member stars.

Table~\ref{catalog} describes the catalog of our open cluster properties. Cross-matching with the published catalogs, we verified the reported and new OCs and provided their OCSN ``Names'' (Cols. 1). The ``OC$\_$flag'' (Cols. 2) was given according to the cross-matching cases, see details in Section~\ref{subsec: crossmatch}. The central coordinates of clusters (Cols. 3-6) were obtained through a two-dimensional Gaussian kernel density estimator (KDE) and the bandwidth of the kernel was calculated via the well-known Scott's rule \citep{Scott1, Scott2}. We derived the mean values of proper motion and parallax of each cluster and their corresponding standard deviations as well (Cols. 7-12). Meanwhile, we fitted the mean radial velocity for each cluster through the Gaussian profile (Cols. 13-16). By visually inspecting the match of the isochrones to the observed cluster CMDs, we further obtained the age, distance modulus, and reddening parameters (Cols. 18-20). In order to better reveal the structural characteristics of nearby star clusters, we provided the radii parameters ($r_c$, $r_t$, $r_o$, $r_e$) of OCSN clusters (Cols. 22-29) according to the two-component model proposed by \citet[hereafter Zhong2022]{2022AJ....164...54Z}. We listed the reported cluster names in the literature as well as the corresponding reference work (Cols. 30-31).

Table~\ref{member cat} describes the catalog of cluster members, including the astrometric and photometric parameters from the {\it Gaia} DR3 (Cols. 1-20), the derived membership probabilities through pyUPMASK (Cols. 21), and the corresponding cluster names in this work (Cols. 22).

\begin{table*}[h]
\centering
\caption{Description of the catalog of open cluster properties.}
\label{catalog}
\begin{tabular}{llcl}
\hline
Column &  Format  & Unit   & Description \\
\hline
Name & string & - & Cluster name in this work\\
OC$\_$flag & string & - & Cluster cross-match cases \\
glon & float & deg & Mean galactic longitude of members \\
glat & float & deg & Mean galactic latitude of members \\
ra & float & deg & Mean right ascension of members \\
dec & float & deg & Mean declination of members \\
pmra & float & mas yr$^{-1}$ & Mean proper motion in right ascension of members \\
e$\_$pmra & float & mas yr$^{-1}$ & Standard deviation of proper motion in right ascension \\
pmdec & float & mas yr$^{-1}$ & Mean proper motion in declination of members \\
e$\_$pmdec & float & mas yr$^{-1}$ & Standard deviation of proper motion in declination \\
plx & float & mas & Mean parallax of members \\
e$\_$plx & float & mas & Standard deviation of parallax \\
RV & float & km s$^{-1}$ & Mean radial velocity of members \\
e$\_$RV & float & km s$^{-1}$ & Standard deviation of radial velocity \\
RV$\_$Flag & string & - & Label of RV \\
N$\_$RV & int & - & Number of RV members \\
N & int & - & Number of members with membership probabilities higher than 0.5 \\
m-M & float & mag & Cluster distance modulus determined by the isochrone fit \\
logt & float & - & Cluster age determined by the isochrone fit \\
E(B-V) & float & mag & Cluster reddening determined by the isochrone fit\\
$r_{\rm{h}}$ & float & deg & Angular size of half number radius\\
$r_{\rm{c}}$ & float & deg & Cluster core radius\\
$e\_r_{\rm{c}}$ & float & deg & Uncertainty of cluster core radius \\
$r_{\rm{t}}$ & float & deg & Cluster tidal radius\\
$e\_r_{\rm{t}}$ & float & deg & Uncertainty of cluster tidal radius\\
$r_{\rm{o}}$ & float & deg & Mean radius of cluster outer region\\
$e\_r_{\rm{o}}$ & float & deg & Uncertainty of the mean radius  of cluster outer region\\
$r_{\rm{e}}$ & float & deg & Cluster boundary radius\\
$e\_r_{\rm{e}}$ & float & deg & Uncertainty of cluster boundary radius\\
Ref\_Name & string & - &  Cluster names in the literature works \\
Ref & string & - & References corresponding to cluster names (see Section~\ref{subsec: crossmatch})  \\
\hline
\end{tabular}
\end{table*}
\begin{table*}[h]
\centering
\caption{Description of the catalog of cluster members.}
\label{member cat}
\begin{tabular}{llcl}
\hline
Column &  Format  & Unit   & Description \\
\hline
Source$\_$id & long & - & Unique source identifier \\
l & double & deg & Galactic longitude \\
b & double & deg & Galactic latitude \\
ra & double & deg & Right ascension (J2016) \\
dec & double & deg & Declination (J2016) \\
plx & double & mas & Parallax\\
plx$\_$err & float & mas & Standard error of parallax \\
pmra & double & mas yr$^{-1}$ & Proper motion in right ascension\\
pmra$\_$err & float & mas yr$^{-1}$ & Standard error of proper motion in right ascension \\
pmdec & double & mas yr$^{-1}$ & Proper motion in declination\\
pmdec$\_$err & float & mas yr$^{-1}$ & Standard error of proper motion in declination\\
rv & float & km s$^{-1}$ & Radial velocity \\
rv$\_$err & float & km s$^{-1}$ & Radial velocity error \\
rv$\_$Flag & string & - & Label of radial velocity members\\
Gmag & float & mag & G magnitude \\
Gmag$\_$err & float & mag & G magnitude error \\
BPmag & float & mag & BP magnitude \\
BPmag$\_$err & float & mag & BP magnitude error \\
RPmag & float & mag & RP magnitude \\
RPmag$\_$err & float & mag & RP magnitude error \\
probs & double & - & Membership probability obtained from pyUPMASK \\
Name & string & - & Corresponding cluster name in this work \\
\hline
\end{tabular}
\end{table*}

\subsection{Radial velocity of OCs}
\label{subsec: rv}

In our cluster member sample, there are a total of 22067 stars with radial velocities from {\it Gaia} DR3 \citep{2022arXiv220605902K}, which can be used to estimate the mean radial velocities (RVs) of 324 clusters. At first, we found that the RVs of the initial cluster members deviate greatly from the median value, especially at $G > 14.5$~mag. It is noticed that a similar situation was also reported by \cite{2022arXiv220714229Y}. To derive the reliable mean radial velocity of each cluster, we removed the outliers beyond 3$\sigma$ and used the high-quality members for calculation. The RV outliers, high-quality members and non-RV members are labeled as ``rv$\_$Flag'' = 0, 1, 2 respectively in the member catalog (see Table~\ref{member cat}). In our sample, the average RVs were derived by the Gaussian fitting for clusters with enough RVs members or simply provided by median RVs for clusters with few RVs members, while the corresponding uncertainties were the 1$\sigma$ of the Gaussian functions or standard deviations, respectively. We provided the corresponding ``RV$\_$Flag'' = 1 for mean RVs of Gaussian-fitting and 2 for median RVs. Eventually, there are 324 OCs with observed mean or median radial velocity (hereafter {\it Gaia} DR3 RVs) which are listed in Table~\ref{catalog}. 

\begin{figure*}[h]
   \centering
   \includegraphics[width=0.8\textwidth, angle=0]{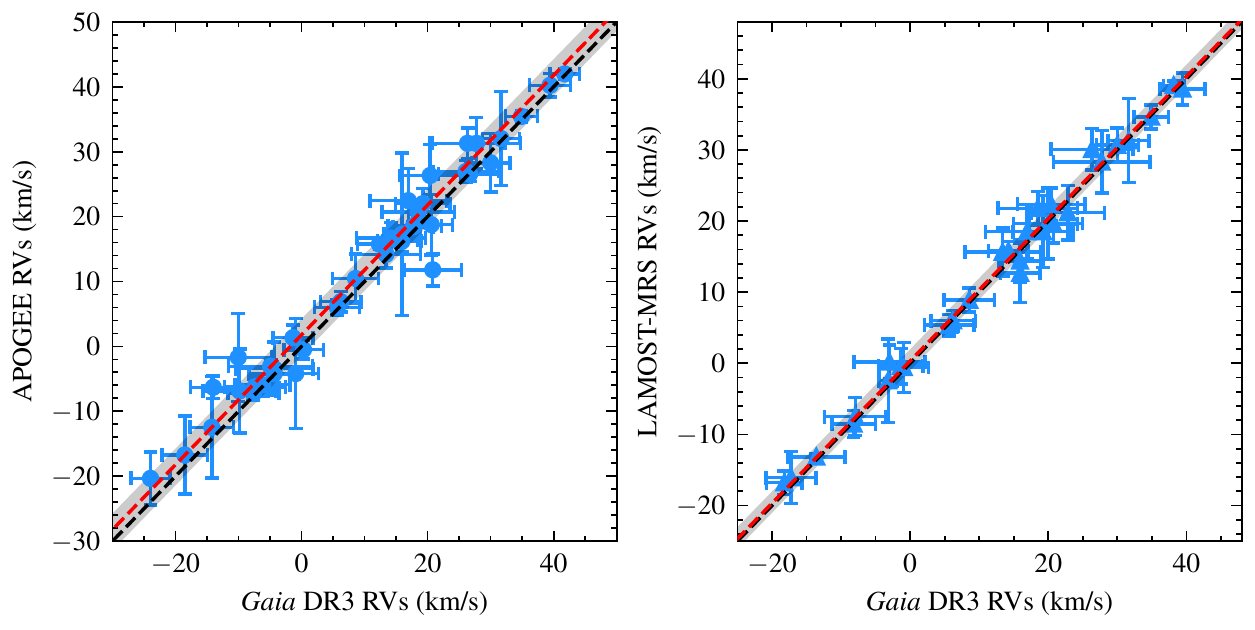}
   \caption{The open cluster mean RVs comparison between {\it Gaia} DR3 and APOGEE/LAMOST-MRS. The error bars of blue points/triangles are the 1$\sigma$ of the Gaussian fitting or standard deviations for open cluster RVs. The black dashed lines are 1:1 scale lines. The red dashed lines are the mean values of the differences between {\it Gaia} DR3 RVs and APOGEE RVs or LAMOST-MRS RVs, and the grey-filled regions are the corresponding 1$\sigma$ regions.}  
   \label{rv comparison}
\end{figure*}

At the same time, we calculated the mean or median RVs of our cluster samples with APOGEE DR17 \citep{2022ApJS..259...35A} and LAMOST DR9 of Medium Resolution survey \footnote{\url{http://www.lamost.org/dr9/}} (hereafter APOGEE RVs and LAMOST-MRS RVs) through the same approach. The comparisons of the RVs for 55 common OCs between {\it Gaia} DR3 and APOGEE, 42 common OCs between {\it Gaia} DR3 and LAMOST-MRS are shown in Figure~\ref{rv comparison}. The mean and standard deviation values of the differences between {\it Gaia} DR3 RVs and APOGEE RVs are -1.42~km s$^{-1}$ and 2.83~km s$^{-1}$, and the corresponding values of the differences between {\it Gaia} DR3 RVs and LAMOST-MRS RVs are -0.48~km s$^{-1}$ and 1.50~km s$^{-1}$, demonstrated as the red dashed lines and the grey-filled regions in Figure~\ref{rv comparison}. It is evident that {\it Gaia} DR3 RVs, APOGEE RVs and LAMOST-MRS RVs are mostly consistent.

\subsection{Metallicity and Isochrone fitting}
\label{subsec: iso}

Before getting the age parameters for our OC samples, we also collected available [Fe/H] metallicity for  clusters from literature spectroscopic work. The Open Cluster Chemical Abundances and Mapping (OCCAM) survey \citep[hereafter Donor2020]{2020AJ....159..199D} provided the [Fe/H] abundances for a sample of 128 OCs from the APOGEE DR16. After cross-matching our cluster samples with Donor2020, 14 common clusters were found. We also acquired 7 common clusters within 500~pc from \citep[hereafter Netopil2022]{2022MNRAS.509..421N}. In addition, we gathered the LAMOST spectroscopic parameters of the other 34 OCs from \citet[hereafter Zhong2020]{2020A&A...640A.127Z}. To ensure the reliability of [Fe/H] for OCs, we select 31 OCs with more than 5 [Fe/H] members, as shown in Table~\ref{metallicity}, which would be used in the following isochrone fitting process. 

\begin{deluxetable}{ccrccc}[h]
\caption{Metallicity of some reported clusters. \label{metallicity}}
\tabletypesize{\footnotesize}
\tablewidth{0pt}
\tablehead{\colhead{Name} & \colhead{Ref\_Name} & \colhead{[Fe/H]} &
\colhead{e$\_$[Fe/H]} & \colhead{N$\_$[Fe/H]} & \colhead{Ref} \\
& & (dex) & (dex) & &
}
\startdata
OCSN\_127 & ASCC$\_$16      & $-$0.06  & 0.06 & 33 & Donor2020 \\
OCSN\_128 & ASCC$\_$19      & $-$0.07  & 0.05 & 19 & Donor2020 \\
OCSN\_129 & ASCC$\_$21      & $-$0.13  & 0.04 & 10 & Donor2020 \\
OCSN\_130 & ASCC$\_$41      & $-$0.11  & 0.07 & 8  & Zhong2020 \\
OCSN\_141 & Alessi$\_$20    & $+$0.14  & 0.04 & 11 & Zhong2020 \\
OCSN\_189 & Collinder$\_$69 & $-$0.10  & 0.05 & 55 & Donor2020 \\
OCSN\_192 & Collinder$\_$350& $-$0.10  & 0.11 & 31 & Zhong2020 \\
OCSN\_194 & Gulliver$\_$6   & $-$0.13  & 0.20 & 18 & Zhong2020 \\
OCSN\_203 & IC$\_$348       & $-$0.17  & 0.14 & 11 & Zhong2020 \\
OCSN\_204 & IC$\_$2391      & $-$0.03  & 0.04 & 11 & Netopil2022  \\
OCSN\_205 & IC$\_$2602      & $-$0.02  & 0.02 & 7  & Netopil2022  \\
OCSN\_206 & IC$\_$4665      & $-$0.01  & 0.02 & 11 & Netopil2022  \\
OCSN\_207 & IC$\_$4756      & $+$0.02  & 0.04 & 13 & Netopil2022  \\ 
OCSN\_213 & L$\_$1641S      & $-$0.09  & 0.07 & 22 & Donor2020 \\
OCSN\_218 & Melotte$\_$20   & $+$0.01  & 0.05 & 64 & Donor2020 \\ 
OCSN\_219 & Melotte$\_$22   & 0.00     & 0.05 & 83 & Donor2020 \\
OCSN\_220 & Melotte$\_$25   & $+$0.12  & 0.04 & 48 & Netopil2022  \\
OCSN\_221 & NGC$\_$752      & $-$0.08  & 0.07 & 49 & Zhong2020 \\
OCSN\_222 & NGC$\_$1039     & $+$0.02  & 0.06 & 7  & Netopil2022  \\
OCSN\_224 & NGC$\_$1662     & $-$0.19  & 0.09 & 35 & Zhong2020 \\
OCSN\_226 & NGC$\_$1980     & $-$0.08  & 0.04 & 9  & Donor2020 \\
OCSN\_227 & NGC$\_$2232     & $-$0.09  & 0.09 & 6  & Zhong2020 \\
OCSN\_228 & NGC$\_$2281     & $-$0.08  & 0.11 & 73 & Zhong2020 \\
OCSN\_234 & NGC$\_$2632     & $+$0.16  & 0.07 & 21 & Netopil2022  \\
OCSN\_241 & NGC$\_$6633     & $-$0.10  & 0.05 & 8  & Zhong2020 \\ 
OCSN\_255 & RSG$\_$1        & $-$0.01  & 0.09 & 20 & Zhong2020 \\
OCSN\_256 & RSG$\_$5        & $+$0.07  & 0.09 & 13 & Zhong2020 \\
OCSN\_259 & Roslund$\_$6    & $-$0.01  & 0.10 & 32 & Zhong2020 \\
OCSN\_261 & Ruprecht$\_$147 & $+$0.12  & 0.03 & 33 & Donor2020 \\
OCSN\_265 & Stock$\_$2      & $-$0.11  & 0.07 & 19 & Zhong2020 \\
OCSN\_266 & Stock$\_$10     & $-$0.13  & 0.09 & 23 & Zhong2020 \\
\enddata
\end{deluxetable}

\begin{figure*}[h]
   \centering
   \includegraphics[width=\textwidth, angle=0]{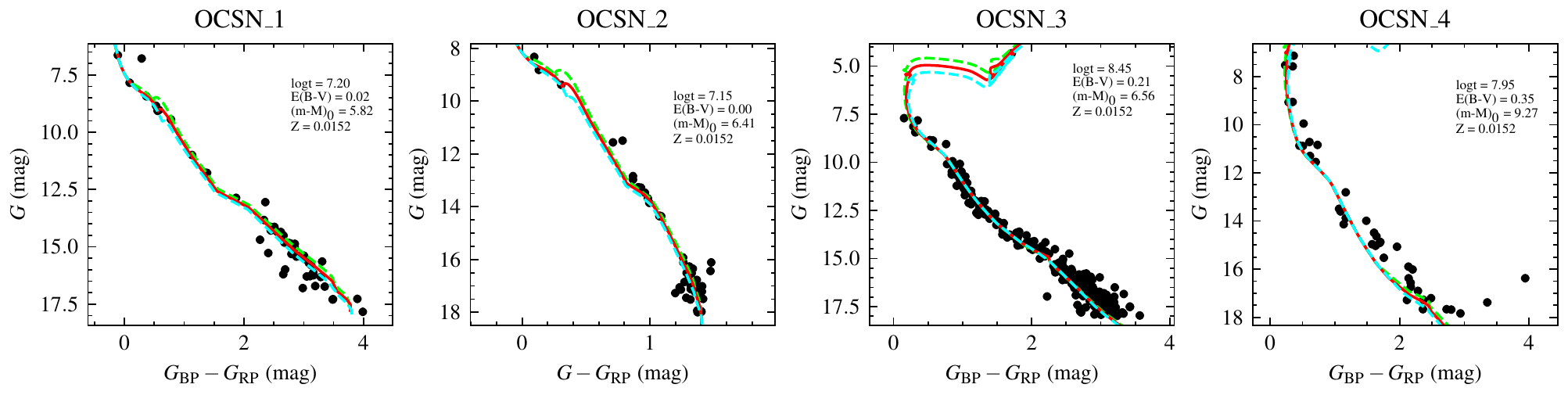}
   \caption{Example of isochrone-fitting results for four clusters: OCSN$\_$1, OCSN$\_$2, OCSN$\_$3, OCSN$\_$4. The black dots refer to the members identified in this work. The red solid lines indicate the best-fitting isochrones while the lime or cyan dashed lines denote the ischrones whose ages (logt) are 0.5 larger or smaller than the best-fitting isochrones. All the isochrones applied the same solar metallicity ($Z_{\odot}$ = 0.0152). The best fitting parameters are shown in the panels for each cluster.}
   \label{iso-fit}  
\end{figure*}

To determine the age parameter of the OCs found in the solar neighborhood, we use a set of Padova isochrones \citep{2017ApJ...835...77M} to perform the CMD fitting. The grid of logarithm ages in isochrones is from 6.0 to 10.10 with an interval of 0.05, and the photometric system is the {\it Gaia} photometric system \citep{2021yCat..36490003R} from CMD 3.6\footnote{\url{http://stev.oapd.inaf.it/cgi-bin/cmd}}. For clusters whose metallicity is reported by literature, we adopted the abundances in Table~\ref{metallicity} as an input parameter to derive the Padova isochrone, while the isochrones of other clusters were adopted with the solar metallicity $Z_{\odot}$ = 0.0152 \citep{2009A&A...498..877C,2011SoPh..268..255C}. We carefully inspected the match of the isochrones to the significant characteristic regions, such as the upper main sequence, the turn-off point, and the red giant or red clump features in the CMDs. By adjusting the isochrones to achieve the best fitting of cluster members in the CMDs, we obtained the age, distance modulus as well as reddening of each cluster. Then, we use the formula $A_{\rm{G}} = 2.74 \times E(B-V)$, $E(BP-RP)=1.339 \times E(B-V)$ and $E(G-RP)=0.705 \times E(B-V)$ \citep{2018MNRAS.479L.102C,2019A&A...624A..34Z} to calculate the E(B-V) values. Figure~\ref{iso-fit} shows the isochrone-fitting examples for four clusters\footnote{The complete isochrone-fitting figure set (324 clusters) is available in the online Journal.} and the final fitting results (age: logt; distance modulus: m-M; reddening value: E(B-V)) are shown in the Table~\ref{catalog}.

\subsection{Structural parameters}

It is evident that the discoverable spatial scale of OCs is greatly expanded in the Gaia era. More and more members located in the extended region were identified through their kinematic properties \citep{2019A&A...621L...3M,2021A&A...645A..84M}. Our investigation of nearby clusters also shows that many clusters have an extended outer structure. Many reported clusters are just tight core components in our OCSN catalog (See Section~\ref{subsec: crossmatch}). 

To better describe the radial density profile (RDP) of clusters with extended outer regions, it is proposed by Zhong2022 of using a two-component model instead of only using a King model \citep{1962AJ.....67..471K,1966AJ.....71..276K}. After deriving the RDP of each cluster through a two-dimensional Gaussian KDE on the spatial space, we further attempted the two-component model to fit the RDP:
\begin{equation}
    F (r) = f (r) + g (r)
\label{comb}
\end{equation}
where $f(r)$ is the King model that mainly described the RDP of core members and $g(r)$ is a logarithmic Gaussian function that described the RDP of corona members \citep{2022AJ....164...54Z}.

In the fitting procedure, the two-component model performs a more reliable approximation of the RDP of most OCSN clusters. In particular, we noted that there are about 33\% of OCSN clusters whose RDP can be well approximated by the single King model. The fraction of clusters that well follow a single King profile is larger than the fraction ( about 10\%) in Zhong2022. We speculate that it is because the outer extended structure of some nearby clusters can be extended to dozens of degrees, the OCSN clusters we identified may still be the core components. However, there are still a fraction of star clusters (43 of 324) that cannot be decently fitted by the single-component or two-component method, possibly due to their sparse distribution or extended tail-like structures or even multiple cores. And it is quite obviously that not all of the OCSN clusters have a clear core.

\section{Discussion}
\label{sec: discussion}

\subsection{Comparison with reported clusters}
\label{subsec: crossmatch}
Based on {\it Gaia} data, several OC-hunting studies have published more than 3000 OCs by applying multifarious clustering algorithms or manually searching approaches. We collected the published OCs within 500~pc from previous works \citep[hereafter LP19, Sim19, CG20, HR21, He22a, He22b, Li22 respectively]{2019ApJS..245...32L,sim_207_2019,2020A&A...640A...1C,2021A&A...646A.104H,2022ApJS..260....8H,2022ApJS..262....7H,2022ApJS..259...19L}, including about 10\% of the OCs within 500~pc of the solar neighborhood. Many nearby young associations and moving groups within 500~pc to the Sun also have been investigated with {\it Gaia} data, such as the Orion complex with the strong sign of radial expansion attributed to a supernova expansion \citep[hereafter K18]{2018AJ....156...84K}, Vela OB2 hosts complex spatial filamentary substructures \citep[hereafter CG19, B20, Pang21]{2019A&A...626A..17C,2020MNRAS.491.2205B,2021ApJ...923...20P}, Taurus region consists of 22 groups \citep[hereafter Liu21]{2021ApJS..254...20L}, Chamaeleon I with two sub-clusters \citep[hereafter R18]{2018A&A...617L...4R}, Corona Australis with ``off-cloud'' and ``on-cloud'' populations \citep[hereafter G20a]{2020A&A...634A..98G}, $\epsilon$ Cha Association \citep[hereafter DV21]{2021AJ....161...87D}, $\rho$ Ophiuchi with two young populations \citep[hereafter G21]{2021A&A...652A...2G}, Perseus with five clustered group Autochthe, Alcaeus, Mestor, Electryon and Heleus \citep[hereafter P21]{2021MNRAS.503.3232P}, stellar `snake' \citep[hereafter T20, W22]{2020ApJ...904..196T,2022MNRAS.513..503W}, Lupus association \citep[hereafter G20b]{2020A&A...643A.148G} and etc. We combined these published OCs as well as  young associations and moving groups related to giant molecular clouds to create a reference cluster catalog (hereafter ref\_OCs). 

In order to carefully analyze the differences between the OCSN catalog and the ref\_OCs catalog, we created a common sample for comparison. However, it is not appropriate to simply cross-match with two cluster catalogs through their center celestial coordinates. This is because the search area in our work (OCSN catalog) is systematically larger than many previous works (ref\_OCs catalog), and the deviation of the cluster center coordinates derived by different methods in different works may be very large. Hence, we adopted cross-matching with all cluster members rather than the cluster itself. Finally, with a matching radius of 1", a common catalog of about 332,000 members was obtained, which is referred to as the common\_memb catalog.

\begin{figure*}[h]
   \centering
   \includegraphics[width=\textwidth, angle=0]{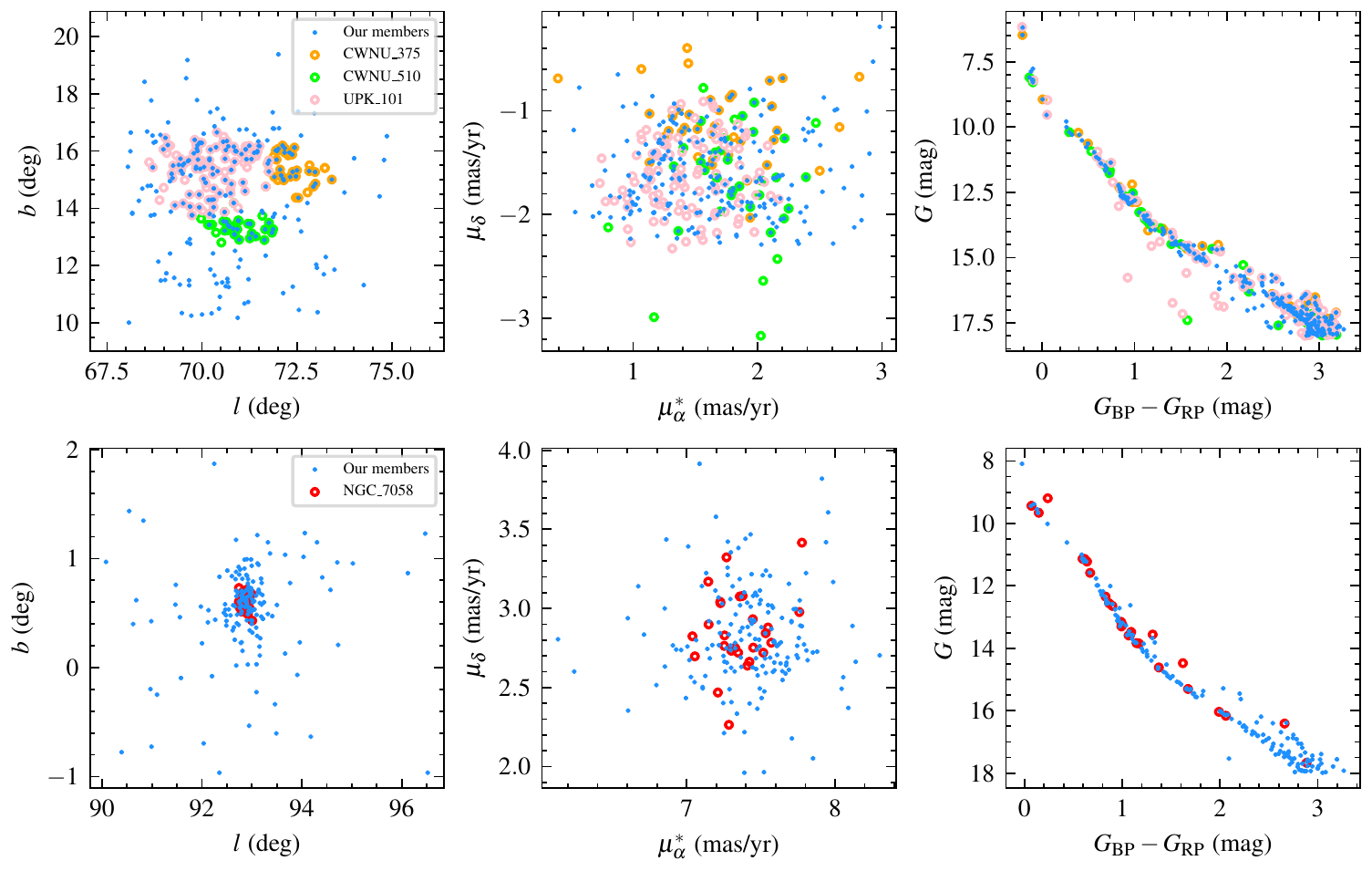}
   \caption{Example of the results of our searching procedure for two clusters. For each cluster, the three scatter plots (from left to right) represent the galactic coordinate distribution, proper motion distribution and CMD. The blue dots refer to the members identified in this work. The orange, green, pink, and red circles represent the members of CWNU$\_$375 and CWNU$\_$510 from He22a, UPK$\_$101 from Sim19, NGC$\_$7058 from CG20 respectively. CWNU$\_$375 and UPK$\_$101 have 4 common members, and CWNU$\_$510 and UPK$\_$101 have 3 common members in previous results.}
   \label{cross}  
\end{figure*}

For each cluster, we calculated the fraction (f) between the common member stars to members in the OCSN catalog and then assigned a flag according to this fraction. Meanwhile, we checked the position, proper motion, and parallax distribution of those common members through visual inspection to assess whether they are the same clusters. In our catalog, we provided the corresponding OC\_flag of the cluster as well as its literature name in Table~\ref{catalog}, which mainly have three cases below:
\begin{enumerate}
 \item [(1)] OC$\_$flag = 1: f $=$ 0\%, the new OCs in the OCSN catalog. 
 \item [(2)] OC$\_$flag = 2: f $<$ 50\% and most of the common members are located on the outer part of the cluster in the OCSN catalog.  
 \item [(3)] OC$\_$flag = 3: f $>$ 50\% or most of the common members are located in the center part of the cluster in the OCSN catalog.
\end{enumerate}

For clusters whose f $<$ 50\%, we further inspected the 5-dimension distribution of their members. For instance, the top panels in Figure~\ref{cross} show the spatial, proper motion, and color-magnitude distributions of members of the three reported clusters. The reported cluster called $\rm{CWNU\_375}$, $\rm{CWNU\_510}$ are both published in He22a, and the $\rm{UPK\_101}$ are published in Sim19. The space distribution clearly shows that the cluster identified in the OCSN catalog incorporates the three reported clusters. In this case, the three reported clusters are regarded as reference clusters that combined as one cluster with $\rm{OC\_flag}$ = 2 in the OCSN catalog (see Table~\ref{catalog}). Similarly, the bottom panels in Figure~\ref{cross} show another cluster case with $\rm{OC\_flag}$ = 3. For this cluster, our work identified more members in a wider range, while the previously reported members of the cluster ($\rm{NGC\_7058}$) in CG20 are only a core component of this cluster. Finally, 25 OCs with OC$\_$Flag = 2 and 198 OCs with OC$\_$Flag = 3 were reported in the previous catalogs, and 101 new OCs with OC$\_$Flag = 1 were not presented in any literature studies. We noticed that about 10\% of our cataloged clusters coincided with known associations and moving groups, which clearly indicates that not all of the clusters are bound open clusters.

\begin{figure*}[h]
   \centering
   \includegraphics[width=\textwidth, angle=0]{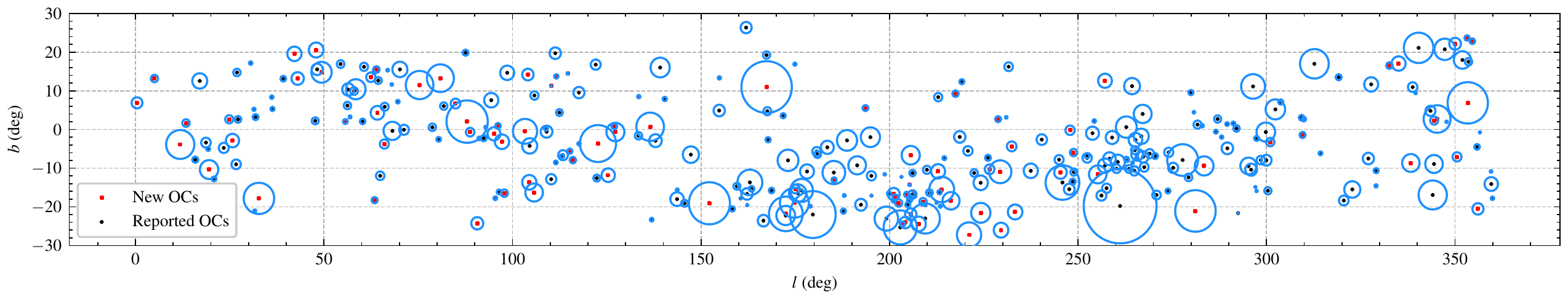}
   \caption{Distribution of the OC population in Galactic coordinates. The black points represent the reported OCs found in this study, and the red squares represent the OCs newly identified in this work using Gaia DR3. The blue circles refer to the $r_{\rm{h}}$ given by this work.}  
   \label{position}
\end{figure*}

Figure~\ref{position} shows the distribution of 324 OCSN clusters in the Galactic coordinates, while the blue circle represents the half-number radius ($r_{\rm{h}}$) of each cluster. Furthermore, we use the red squares and black dots to present the new OCs( OC\_flag = 1) and reported OCs (OC\_flag = 2,3) respectively. It is worth noting that many nearby clusters with large spatial scale ($r_{\rm{h}}$) were discovered for the first time in our work. This also shows that our slicing approach is very effective for searching nearby OCs with a large spatial distribution.

\begin{figure*}[h]
   \centering
   \includegraphics[width=\textwidth, angle=0]{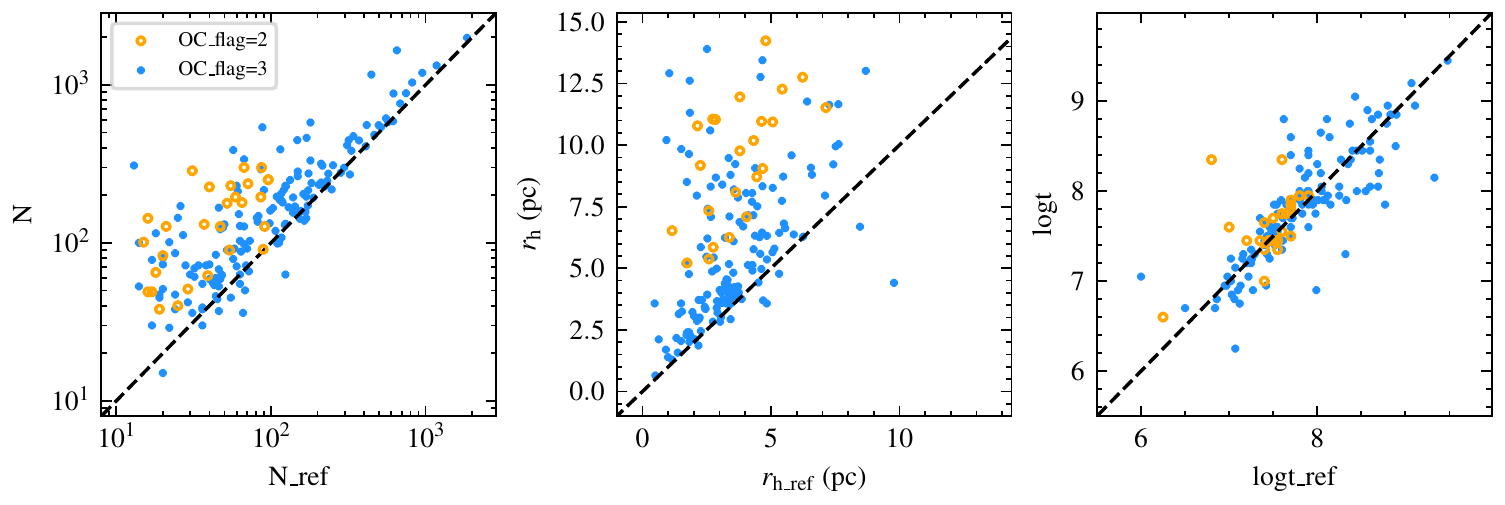}
   \caption{The comparison distribution of member numbers, $r_{\rm{h}}$ and age parameters between the reported clusters from ref\_OCs and the OCs obtained by this work. The blue dots refer to the OCs with ``OC\_flag'' = 3 and the orange circles refer to the OCs with ``OC\_flag'' = 2. The black dashed lines both show the relation of y=x.}  
   \label{parameter com}
\end{figure*}

For OCSN clusters that have been reported, we also compared their number of members, the half-number radius in physical sizes, and the age with the literature results. Figure~\ref{parameter com} shows the comparison results, while the orange circles and blue dots present clusters whose OC\_flag=2 and OC\_flag=3 respectively. It can be seen that our cluster sample (especially the cluster with OC\_flag=2) contains more member stars than the literature results. Moreover, because the $r_{\rm{h}}$ of most clusters is larger than previous results, the updated members in the cluster present a more extended spatial distribution. Our OCSN catalog expands the physical size of many nearby star clusters. On the other hand, because we only increased the number of member stars, the age of isochrone-fitting results of star clusters is still keeping consistent with the literature results.

\subsection{Binary clusters}
\label{subsec: BC}

OCs are born in giant molecular clouds and in some cases also formed in groups \citep{2016MNRAS.455.3126C}. A number of OCs are found in pairs or higher-order systems \citep{1976Ap.....12..204R,1995A&A...302...86S,2019A&A...623C...2S}. As the most famous double star clusters, $h$ and $\chi$ Persei have been extensively studied and many interesting results have been obtained \citep{2002ApJ...576..880S,2019A&A...624A..34Z,2019ApJ...876...65L}. Based on the high precision kinematic information provided by {\it Gaia} data, more and more binary clusters have been confirmed \citep{2019A&A...623C...2S,2021ARep...65..755C,2021MNRAS.503.5929B,2022MNRAS.510.5695A}, which is important for studying the formation and dynamical evolution of OCs \citep{2009A&A...500L..13D,2010ApJ...719..104D,2017MNRAS.471.2498A}. We performed a preliminary screening in our cluster samples with spatial separations $\rm{\Delta pos} < 20~pc$ \citep{1995A&A...302...86S} and velocity differences $\Delta$V $<$ 5~km~ s$^{-1}$ \citep{2019A&A...623C...2S}. As a result, we got 15 groups of OCs including binary and triple cluster systems. The results are listed in Table~\ref{binary clusters}.

\begin{figure*}[h]
   \centering
   \includegraphics[width=\textwidth, angle=0]{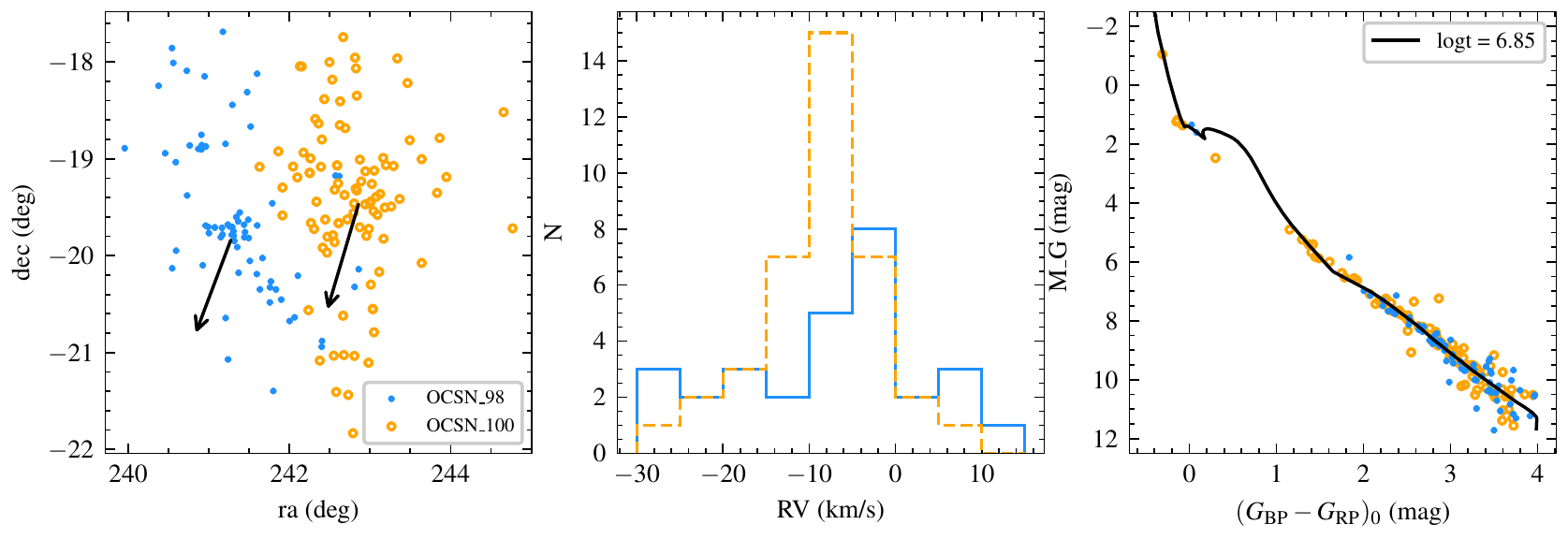}
   \caption{An example of binary cluster in our samples, OCSN$\_$98 in blue and OCSN$\_$100 in orange. Left panel: the spatial distribution for two OCs and the black arrows refer to their projected moving direction. Middle panel: the histograms of radial velocities for their members.  Right panel: the absolute CMDs for two OCs and the black line refers to the isochrone with solar metallicity $Z_{\odot}$ = 0.0152 \citep{2009A&A...498..877C,2011SoPh..268..255C} and logarithm of age logt $=$ 6.85.}
   \label{OCSN98_100}
\end{figure*}

In our catalog, there are 19 open cluster pairs with a common origin, whose age differences are less than 30~Myr. We show a pair of two new OCs in Figure~\ref{OCSN98_100} : OCSN\_98 and OCSN\_100. It is clear that the mean central position of the two clusters is very close ($\Delta$pos $=$ 13~pc), while the tangential velocities (black arrows in the left panel) and radial velocities (histogram in the middle panel) are also similar. The total velocity difference between the two clusters is $\Delta$V $=$ 1.3~km~ s$^{-1}$. The absolute CMDs of the two clusters are presented in the right panel, which presents the same visual fitting age with logt = 6.85. The similarity of the two clusters suggests that they may have a common origin.

Furthermore, along with more new star clusters added, we found three groups (Group 3, Group 5, and Group 11 in Table~\ref{binary clusters}) containing triple OCs, whose age differences are less than 10~Myr. For example, Group 3 comprises two new OCs and one already reported open cluster: OCSN\_40, OCSN\_41, and OCSN\_158. As shown in Table~\ref{binary clusters}, the two pairs of OCSN\_40, OCDN\_41 and OCSN\_40, OCSN\_158 have similar positions ($\Delta$pos $\approx$ 17pc) and similar velocities($\Delta$V $ < $ 3.5 km~ s$^{-1}$). At the same time, since these three clusters have almost the same age (logt $=$ 7.25, 7.20, 7.20), it can be inferred that the triple clusters also formed together from the same molecular cloud. 

\begin{deluxetable}{ccccc}[h]
\caption{Binary clusters in our OC sample. \label{binary clusters}}
\tabletypesize{\normalsize}
\tablehead{\colhead{Group} & \colhead{cluster1} & \colhead{cluster2} & \colhead{$\Delta$pos} & \colhead{$\Delta$V}\\
& & & (pc) & (km~  s$^{-1}$) 
}  
\startdata
1 & OCSN\_16 & OCSN\_18 & 15.9 & 1.4 \\
2 & OCSN\_29 & OCSN\_286 & 16.8 & 2.8 \\
3 & OCSN\_40 & OCSN\_41 & 17.1 & 1.0 \\
3 & OCSN\_40 & OCSN\_158 & 17.5 & 3.5 \\
4 & OCSN\_50 & OCSN\_51 & 12.3 & 3.4 \\
5 & OCSN\_91 & OCSN\_92 & 6.5 & 3.6 \\
5 & OCSN\_91 & OCSN\_237 & 19.9 & 1.6 \\
5 & OCSN\_92 & OCSN\_237 & 17.1 & 2.3 \\
6 & OCSN\_98 & OCSN\_100 & 13.1 & 1.3 \\
7 & OCSN\_118 & OCSN\_271 & 16.5 & 1.0 \\
8 & OCSN\_124 & OCSN\_184 & 16.8 & 3.8 \\
9 & OCSN\_127 & OCSN\_129 & 13.2 & 1.7 \\
10 & OCSN\_128 & OCSN\_283 & 18.1 & 2.4 \\
11 & OCSN\_176 & OCSN\_ 178& 18.9 & 0.8 \\
11 & OCSN\_177 & OCSN\_178 & 12.7 & 2.1 \\
12 & OCSN\_187 & OCSN\_188 & 5.6 & 4.3 \\
13 & OCSN\_197 & OCSN\_198 & 18.2 & 1.4 \\
14 & OCSN\_245 & OCSN\_246 & 10.0 & 3.5 \\
15 & OCSN\_322 & OCSN\_324 & 10.9 & 2.4 \\
\enddata
\end{deluxetable}

\section{Summary}
\label{sec: sum}
In this paper, we performed a systematically blind search for OCs at Galactic latitudes $|b| \le 30 ^{\circ}$ within 500~pc of the solar neighborhood by choosing different slicing box sizes in different distance grids with {\it Gaia} DR3 data. By utilizing the clustering algorithms pyUPMASK and HDSBSCAN, we identified a total of 324 OCs. Our results include 101 new clusters that were never reported before, increasing the OC census within 500~pc by about 50\%. Meanwhile, 223 reported clusters and their members were updated by carefully comparing the spatial distribution and other properties with the previous cluster catalog (ref\_OCs). In the OCSN catalog, we provided the membership probabilities of member stars and further estimated the mean positions, proper motions, parallaxes, and structural parameters for each cluster. We also derived mean radial velocities of OCs through the Gaussian fitting based on {\it Gaia} DR3. Subsequently, we performed the visual isochrone-fitting to obtain the ages, distance modulus, and reddening values for the clusters according to the distribution of member stars on the CMDs.

In particular, we compared the star clusters in the literature with our star clusters and use an OC\_flag to classify the OCSN clusters into three samples. Our classification based on manual inspection not only marks new clusters but also combines some duplicate or partially reported clusters. Additionally, 19 pairs of clusters were identified as binary clusters in the solar neighborhood, and 3 groups of OCs were confirmed as triple cluster systems, with spatial separation less than 20 pc, velocity difference less than 5 km s$^{-1}$, and similar ages.

For our hunted OC samples within 500~pc in the solar neighborhood, more detailed analyses are needed to further investigate their properties, such as the mass function and the dynamical states. Especially more spectroscopic data for the member stars will be of prime importance to determine the dynamical and chemical evolution of these clusters.

\begin{acknowledgements}
This work is supported by the National Key R\&D Program of China No. 2019YFA0405501. 
Li Chen acknowledges the support from the National Natural Science Foundation of China (NSFC) through grants 12090040 and 12090042. 
Jing Zhong would like to acknowledge the NSFC under grant 12073060, and the Youth Innovation Promotion Association CAS. 
We acknowledge the science research grants from the China Manned Space Project with NO. CMS-CSST-2021-A08.

This work has made use of data from the European Space Agency (ESA) mission {\it Gaia} (\url{https://www.cosmos.esa.int/gaia}), processed by the {\it Gaia} Data Processing and Analysis Consortium (DPAC, \url{https://www.cosmos.esa.int/web/gaia/dpac/consortium}). Funding for the DPAC has been provided by national institutions, in particular, the institutions participating in the {\it Gaia} Multilateral Agreement.

\end{acknowledgements}
\bibliography{ms}
\end{CJK*}
\end{document}